**Perspective**

Responding to an enquiry concerning the geographic population structure (GPS) approach and the origin of Ashkenazic Jews – a reply to Flegontov et al.

Ranajit Das[1], Paul Wexler[2], Mehdi Pirooznia[3], and Eran Elhaik[4*]

[1] Manipal Centre for Natural Sciences (MCNS), Manipal University, Manipal, Karnataka, India
[2] Tel Aviv University, Department of Linguistics, Tel-Aviv, Israel 69978
[3] Johns Hopkins University, Department of Psychiatry and Behavioral Sciences, Baltimore, MD, USA 21205
[4] University of Sheffield, Department of Animal and Plant Sciences, Sheffield, UK S10 2TN

* Please address all correspondence to Eran Elhaik at e.elhaik@sheffield.ac.uk

**Abstract**

Recently, we investigated the geographical origins of Ashkenazic Jews (AJs) and their native language Yiddish by applying a biogeographical tool, the Geographic Population Structure (GPS), to a cohort of 367 exclusively Yiddish-speaking and multilingual AJs genotyped on the Genochip microarray. GPS localized most AJs along major ancient trade routes in northeastern Turkey adjacent to primeval villages with names that may be derived from the word "Ashkenaz." These findings were compatible with the hypothesis of an Irano-Turko-Slavic origin for AJs and a Slavic origin for Yiddish and at odds with the Rhineland hypothesis advocating a German origin of both. Our approach has been adopted by Flegontov et al. (2016a) to trace the origin of the Siberian Ket people and their language. Recently, Flegontov et al. (2016b) have raised several questions concerning the accuracy of the Genochip microarray and GPS, specifically in relation to AJs and Yiddish. The authors have also questioned basic elements in the theory of the evolution of languages. Although many of these issues have been addressed in our previous papers, we take this opportunity to clarify the principles of the GPS approach, review the recent biogeographical and ancient DNA findings regarding AJs, and comment on the origin of Yiddish.

**Background**

Recently, Flegontov and colleagues (2016b) published an enquiry concerning the accuracy of the GenoChip microarray (Elhaik et al. 2013) and the Geographic Population Structure (GPS) tool (Elhaik et al. 2014), which infers the geographic origin of individuals provided their genotype data. The authors have also questioned the geographical and ancestral origins of Ashkenazic Jews (AJs) and their language in light of three biogeographical analyses (Behar et al. 2013; Elhaik 2013; Das et al. 2016).

Since the growing usage of GPS to study deep origins of populations and languages necessitates elaborating the strengths and limitations of this framework, we provide here point by point answers to the questions posed by Flegontov et al. (2016b). We show that published biogeographic analyses are consistent with a Turkish origin for AJs and briefly discuss the question of Yiddish origins. The discourse is summarized in Table 2. We note with interest that prior to their current enquiry, Flegontov et al. (2016a) adopted the Genochip microarray and GPS tool to find the origin of the Siberian Ket people, considered the last nomadic hunter-gatherers of Siberia whose language has no apparent affiliation with any language family. We are glad that this has led the authors to question these technologies and will use this opportunity to address their concerns.

Understanding the GPS framework (questions #1-9)

Over 135 years ago, Alfred Russel Wallace (1878) first speculated on the global biodiversity patterns in what became the core mission of biogeography: explaining the geographical spatial patterns of global biodiversity and exploring their implications (Gaston 2000). However, this was not a new challenge for human biogeographers. Scientists have been searching for a method that allows tracing humans to their geographical origins since the time of Herodotus of Halicarnassus (Rowe 1965), yet only in the past decade were high-throughput genetic data harnessed to answer this question. Existing biogeographical approaches have been applied to identify the geographical origin of modern-day individuals down to the level of linguistic boundaries (Barbujani and Sokal 1990) and neighboring countries (Novembre et al. 2008), but localization of worldwide individuals to countries remains a significant challenge (Elhaik et al. 2014).

Elhaik, Tatarinova, and colleagues (2014) proposed a new paradigm for the problem of human biogeography, termed Geographic Population Structure (GPS). Dismissing ethnic notions, the GPS framework assumes that all humans are admixed and that their genetic variation or admixture can be modeled by the proportion of genotypes assigned to regional gene pools. Building on the work of Cavalli-Sforza and other investigators (Cavalli-Sforza, Menozzi, and Piazza 1994; Eller 1999; Relethford 2001) who established a strong relationship between genetic and geographical distances, GPS infers the geographical coordinates of an individual by matching their admixture proportions with those of reference populations as long as they are known to reside in a certain geographical region for a substantial period of time. Intuitively the reference populations can be thought of as "pulling" the individual in their direction with a strength proportional to their genetic similarity until a consensus is reached (Figure 1), much as a Global Positioning System determines the location of a car using satellites orbiting Earth.

Elhaik and colleagues (2014) tested GPS on four different datasets consisting of over 2,000 individuals analyzing various subsets of Genochip markers (Elhaik et al. 2013) ranging from ~40,000 to ~130,000 in size. GPS's accuracy was evaluated using the leave-one-out procedure at the individual and population levels, with the latter being more stringent. Applied to a worldwide population dataset and using the leave-one-out individual approach, GPS correctly assigned 83% of worldwide individuals to their country of origin, and, when applicable, ~66% of them to their regional location with high sensitivity (0.75) and specificity (0.99). In terms of distances from region of residency, GPS placed 50% of worldwide individuals within 87 kilometers (km) from their region with 80 and 90% of them within 645 and 1,015km from their region, respectively. Applied to over 200 Oceanians, GPS localized 87.5% of the individuals to their home island. Applied to nearly 300 Sardinians, GPS placed a quarter of the individuals to their village, half within 15km, and 90% of individuals within 100km of their home with higher accuracy in high altitude regions characterized by endogamy and relative isolation. Elhaik et al.'s (2014) findings presented GPS as a promising biogeographical tool in terms of its sensitivity and specificity.

Understanding the origin of Ashkenazic Jews and Yiddish (questions #10-18)

The geographical origin of Ashkenazic Jews, Yiddish, and the Biblical "Ashkenaz" are among the longest standing questions in history. The first known discussion of the origin of German Jews and Yiddish surfaced in the writings of the Hebrew grammarian Elia Baxur in the first half of the 16$^{th}$ century (Wexler 1993). "Ashkenaz" is one of the most disputed Biblical placenames, and the debate regarding its accurate location is much older. That placename appears in the Hebrew Bible as the name of one of the descendants of Noah and as a reference to the kingdom of Ashkenaz, prophesied to be called together with Ararat and Minnai to wage war against Babylon (Jeremiah 51:27).

Weinreich (2008), the doyen of the field of modern Yiddish linguistics (1894-1969), emphasized a truism that the history of Yiddish mirrors the history of its speakers. It is well established that this history is also reflected in the DNA through relationships between genetics, geography, and language (Cavalli-Sforza 1997; Kitchen et al. 2009; Balanovsky et al. 2011; Bouckaert et al. 2012). This prompted Das et al. (2016) to infer the biogeographical origin of sole-Yiddish speaking and multilingual AJs using GPS. The findings were evaluated in light of two competing linguistic hypotheses, i.e., the Rhineland and Irano-Turko-Slavic hypotheses [Das et al. (2016), Table 1]. GPS traced nearly all AJs and some Sephardic Jews (Mountain Jews) to major ancient trade routes in northeastern Turkey adjacent to four primeval villages whose names resemble "Ashkenaz:" İşkenaz (or Eşkenaz), Eşkenez (or Eşkens), Aşhanas, and Aschuz. AJs were also found to be genetically closest to Turk, southern Caucasian, and Iranian populations, suggesting a common origin in Iranian "Ashkenaz" lands. These findings were more compatible with an Irano-Turko-Slavic origin for AJs and a Slavic origin for Yiddish than with the Rhineland hypothesis, which lacks historical, genetic, and linguistic support (Das et al. 2016) and relies on fictitious and supernatural elements (Table 1) that have no place in science (van Straten 2003; Elhaik 2013). Our findings have also highlighted the strong social-cultural and genetic bonds of Ashkenazic and Iranian Judaism and their shared Iranian origins.

Question #1: How should GPS predictions be interpreted?
Flegontov et al. (2016b) have questioned the meaning of GPS predictions. Unlike existing biogeographical approaches based on principal component analysis or alike analyses (Novembre et al. 2008; Yang et al. 2012; Elhaik 2013), GPS is an admixture-based approach. Observing the distribution of admixture components in worldwide populations (Elhaik et al. 2014, Figure 1) yields several patterns: First, most populations have characteristic admixture proportions. Second, neighboring populations share similar admixture proportions. Third, admixture components are relatively geographically localized. These observations underlie the principles of the GPS framework, which suggests that given a global admixture network as in Figure 1 of Elhaik et al. (2014) making relative geographical inferences for a test individual with certain admixture proportions is feasible.

GPS analyzes non-coding, non-functional (Graur et al. 2013) ancestry informative markers (AIMs) (Elhaik et al. 2013). AIMs allow identifying populations that vary in substructure or the degree of admixture and detecting subtle population subdivisions (Enoch et al. 2006), which reduces problems of misclassification.

Population structure is affected by biological and demographic events like genetic drift, which acts fast on small, relatively isolated populations and slowly on large, non-isolated populations, and migration, which is more frequent (Elhaik 2012; Jobling, Hurles, and Tyler-Smith 2013). To understand the relationships between geography and the formation of admixture proportions, we should consider the effect that both relative isolation and migration history had on the allele frequencies of populations. Unfortunately, oftentimes we lack information about both processes. The GPS framework addresses this problem by analyzing the relative proportion of admixture in a global network of reference populations that provide us with different "snapshots" of historical admixture events. These global admixture events occurred at different times through different biological and demographic processes and are directly related to our ability to associate an individual with their matching admixture event.

In populations that exhibit high relative isolation due to isolation by distance, the admixture event was old, and GPS can localize the test individual with their parental population more accurately. By contrast, if the admixture "event" was recent and the population did not maintain relative isolation, GPS prediction would be erroneous (Figure 2). This is the case of most North Americans, Israeli Jews (Elhaik 2016), and Caribbean populations (Elhaik et al. 2014), whose admixture proportions still reflect the massive 19$^{th}$ and 20$^{th}$ centuries admixture events involving Native Americans, West Europeans, and Africans. While we still do not know the original level of isolation, we can differentiate these two cases, by examining the similarity between the admixture proportions of the test individual and those of populations from the predicted location. If this similarity is high, we can conclude that we have inferred the likely location of the admixture event that shaped the admixture proportion of the test individual. If the similarity is low, we can conclude that either the individual is mixed or that the parental population does not exist either in GPS's reference panel or in reality. Interestingly, most of the time (83%) GPS predicted unmixed individuals to their true locations with most of the remaining individuals predicted to neighboring countries (Elhaik et al. 2014).

To understand how migration affects the admixture proportions of the migratory and host populations, we can consider three simple cases of point or massive migration followed by assimilation and migration followed by isolation. Point migration events have little effect on the admixture proportions of the hosting population, particularly in the case of a paucity of migrants absorbed by the host population. In such case, the migrants' admixture proportions would resemble those of the host population within a few generations, depending on their initial values. The resting place of the migrants therefore represents the last place that admixture has occurred. Massive demographic changes, such as large-scale invasion or migration, that affect a large part of the population are rarer and create temporal shifts in the admixture proportions of the host population, which will temporarily appear as a two-way mixed population until the admixture proportions 'level off.' This also depends on where the migratory population came from. Here again, the geographical placement of the host population represents the last place that admixture has occurred at the population level for both populations. In both cases, if applied after admixture proportions have 'levelled off,' GPS would predict the location of the host population for both the host and migratory populations. When a population migrates from point A to B and maintains genetic isolation, it distorts the genetic-geographic model (Ramachandran et al. 2005). Such populations are easily detectable since they will not be predicted to their contemporary region of residency in the leave-one-out population approach and be excluded from the reference panel, as previously described (Das et al. 2016). In this case, GPS will predict the migratory population to point A, which is also where the admixture that shaped its admixture proportions took place. While human migrations are not uncommon, maintaining a perfect genetic isolation over a long period of time is very difficult (Veeramah et al. 2011; Behar et al. 2012; e.g., Elhaik 2012; Elhaik 2016; Hellenthal et al. 2016). GPS predictions for the vast majority of worldwide populations indicate that these cases are indeed exceptional (Elhaik et al. 2014; Das et al. 2016).

Question #2: How does GPS behave in the case of two-way mixed individuals?
Flegontov et al. (2016b) next asked how the localization of two-ways mixed individuals should be interpreted? The current GPS version is unsuitable for analyzing two-ways mixed individuals (e.g., Chinese-British) and will report the middle location of the parental populations (in this case, South Russia) since both parental populations are "pulling" in equal strengths. Das et al. (2016) showed that erroneous localization can be easily identified since the admixture proportions of the test individual will largely deviate from those of native individuals (Figure 1). Flegontov (2016a) set a prediction certainty threshold according to which "prediction uncertainty over 4% indicates that the individual is of a mixed origin and the GPS algorithm is not applicable." Therefore, GPS results can also be interpreted as the average coordinates of the individual's ancestors. The same logic also applies to "softly mixed" individuals whose parents are from different villages, assuming that the parental populations are represented in the reference populations and can "pull" the test individual.

Question #3: Is tracing a population movement back in time feasible?
GPS produces a single geographical location (not movements), and it does not have a dating component for the admixture event. Dating the age or ages of the genetic-geographic model is a very challenging task since both admixture (Falush, van Dorp, and Lawson 2016) and dating tools are inadequate (Pugach et al. 2011; e.g., Loh et al. 2013; Sanderson et al. 2015). GPS's prediction cannot predate the time periods when the admixture events that shaped the population structure of the reference populations occurred. Therefore, although GPS identified a 3,000 year old population structure in Oceania, caution is advised in dating the admixture events using external tools or other resources.

Question #4: Does the choice of Genochip markers bias GPS predictions?
Flegontov et al.'s (2016b) questioned the choice of the GenoChip microarray used by Flegontov et al. (2016a) and Das et al. (2016). The authors pondered whether the choice of 100,000 markers may bias GPS's predictions and whether rare alleles or whole-genome data would be preferred over common alleles. In our recent studies, we applied GPS to various subsets of Genochip markers consisting of both common and rare AIMs (Elhaik et al. 2013). Elhaik et al. (2014) have demonstrated that accurate GPS predictions can be made with 40,000 markers, yet a decreasing number of markers bias the analysis. Such may be the case with the recent study of Flegontov et al. (2016a), where GPS was applied on a reduced dataset of ~30,000 markers to study the origin of the Siberian Ket people (Flegontov et al. 2016a). Ignoring the effect of reduced datasets on the performances of GPS is bound to bias its predictions. Unfortunately, whole genome data remain unaffordable and unavailable for most populations.

Question #5: Did results from an older study prompt the localization in a later study?
Flegontov et al.'s (2016b) asked whether Elhaik et al.'s (2014) localization of Sardinians to villages prompted the later localization of Ashkenazic Jews to primeval villages (Das et al. 2016). Flegontov et al.'s (2016b) conflated two different studies. Elhaik et al. (2014) provided a proof-of-concept to the accuracy of GPS by, among else, localizing Sardinians to their villages. In a separate analysis, Das et al. (2016), applied GPS to localize Ashkenazic Jews to ancient Iranian lands in northeastern Turkey in a region that harbored four ancient villages whose names resemble the word "Ashkenaz." Interestingly, the Lesgian people of the Caucasus still call their neighbors, the Iranian-speaking Mountain Jews "Ashkenazim"–the original meaning of which was "Scythians" (Byhan 1926; Wexler 2016). The partial Iranian origin of AJ was further inferred based on the genetic similarity of AJs to Sephardic Mountain Jews and Iranian Jews as well as their similarity to Near Eastern populations and simulated "native" Turkish and Caucasus populations.

There are very good grounds therefore for inferring that those Jews who considered themselves Ashkenazic adopted this name and spoke of their lands as Ashkenaz, since they perceived themselves as of Iranian origin. That we find varied evidence of the knowledge of Iranian language among Moroccan and Andalusian Jews and Karaites prior to the 11[th] century is a compelling point of reference to assess the shared Iranian origins of Sephardic and Ashkenazic Jews (Wexler 1996). It is important to note that Iranian-speaking Jews in the Caucasus (the so-called Juhuris) and Turkic-speaking Jews in the Crimea prior to World War II called themselves "Ashkenazim" (Weinreich 2008). Therefore, our inference is supported by genetic, linguistic, and historical evidence, which we believe has more weight as a simple origin that can be more easily explained compared to a more complex scenario that involved multiple translocations.

Question #6: Was the localization of Sardinians to their villages due to noise?

Flegontov et al. (2016b) next asked whether a difference of 1-2% in the admixture proportions of the Sardinian villagers is due to admixture noise or population structure? It is expected that the admixture differences between adjacent villages would be small since there are no impenetrable barrier between them, however it is easy to see that these differences are not noise. Noise is expected to distribute randomly across all admixture components and this is not the case. Such "leaky" admixture components would result in larger differences between the villages, rather than small ones as Flegontov et al. (2016b) noted.

When there is high localization of gene pools, GPS has very low levels of noise (Elhaik et al. 2014, Figure 1). For example, the proportion of the Northeast Asian component in African is $10^{-5}$. Equal noise levels are found in the Southern African component in East Asian and North European populations, the Native American and Oceanian components in African and Middle Eastern populations, and the Subsaharan African component in Native American populations. Therefore, Flegontov et al.'s (2016b) comment that the "placement of a quarter of Sardinians into their home villages was possible due to these differences in admixture profiles" is correct and is consistent with the work of Flegontov et al. (2016a) who considered an admixture difference of 2.5% valid to classify Ket individuals as Kets.

We agree with Flegontov et al. (2016b) that a "variability of 1-2% in absolute values is generally considered as noise in admixture analyses, and depends much on dataset composition and on the number of algorithm iterations, among which the best one is selected" in an unsupervised admixture setting. However, GPS is based on a supervised admixture setting which is more robust to noise and does not involve multiple iterations (presumably the authors refer here to the choice of *K* subdivisions).

Finally, we wish to correct the impression that GPS aims to predict phonebook addresses from genetic data. GPS converts genetic data into the place where the admixture event that shaped the individual's genome took place. In the absence of perfect knowledge on isolation and demographic history of populations, GPS predictions were validated by using unmixed individuals who claimed descent from certain countries, regions, islands, or villages. Elhaik et al. (2014) analyzed four different datasets and showed that for the vast majority of worldwide individuals, GPS's geographical inferences are within very short distances from the individual's current place of residency.

Question #7: Does absence of reference populations reduce the accuracy of the results?
Since GPS relies on reference populations to gravitate test individuals, a comprehensive coverage of these populations is necessary to derive accurate results, just as global satellite coverage is essential to derive the accurate location of a car. Similarly to that system, the higher the coverage the more accurate the localization. The consequences of gaps in the global coverage would vary based on the location of the gap. An inland gap may result in a small or no error since the surrounding reference populations will contribute to the localization (Figure 1). By contrast, gaps in islands, shores, or simply remote locations would result in inaccurate localization since the test individual would gravitate towards more remote areas. Testing the accuracy of the localization can be done by comparing the admixture proportions of the test individual and the "native" population, as was done in Das et al. (2016). We also refer Flegontov et al. (2016b) to Flegontov et al. (2016a), who addressed this question by correctly noting that GPS reports the smallest distance to the nearest reference population, which can be used to infer the absence of key reference populations. In other words, we can effectively measure the degree that latter admixture shifted the individual away from its original relative isolated population in areas with high coverage of reference populations.

Question #8: Can inaccurate localizations be due to shared ancestry or inaccurate modelling?
Flegontov et al.'s (2016b) suggested that the localization of a few English and Italians to Germany and Greece, respectively, was not due to their shared ancestral origins. We question this suggestion as these populations have a long and well documented history of gene exchange (e.g., Leslie et al. 2015; Fiorito et al. 2016). Incorrect modelling is expected to result in spurious erroneous localizations, which have not been observed in GPS analyses (Elhaik et al. 2014; Das et al. 2016; Flegontov et al. 2016a).

Question #9: Why are GPS predictions incorrect for some of the non-Jewish populations in Das et al. (2016)?
Our methods are unlikely to perform at perfect accuracy particularly under the more restrictive leave-one-out population approach as was used in this trial (Das et al. 2016). In such approach, GPS's accuracy is evaluated if GPS can localize a population (e.g., Germans) to their region of residency (e.g., Germany) solely based on the "pulling" of neighboring populations (e.g., non-Germans). This high bar was intentionally set to allow estimating the expected error when predicting populations that are not represented in the reference panel, as in the case with AJs (see also question #7). As expected (Elhaik et al. 2014), coast-line populations and populations that were not surrounded by related populations (in this reference panel), like Tuscan Italians and Mongols, were predicted with higher error.

The dense central Eurasian reference population panel still allowed GPS to assign 83% and 78% of the individuals to within 500km or 250km from the political boundaries of their country or regional locations despite the restrictions imposed by the leave-one-out population approach. Individuals who speak geographically localized languages were predicted with nearly perfect accuracy (97% and 94% of the individuals were assigned within less than 500km and 250km of their countries, respectively). On average, AJs were predicted to within 211km from at least one of the primeval villages once existed in in northeastern Turkey, consistent with the margins of error obtained for the general population.

Question #10: Can we infer the likelihood of European admixture in AJs from the demographic data?
Flegontov et al.'s (2016b) reasoned that since 86% of AJs "originated from the USA" a "recent European admixture in these Jewish samples is rather likely." It is correct that USA is the modern-day residency [Das et al. (2016), Table 2] of the AJs in our study and that most of the AJs have European origins [Das et al. (2016), Figure 4], however it is impossible to infer the likelihood of European admixture from this information. Eighty percent of the AJs reported having four AJ grandparents, and their results were very similar to those of the remaining cohort (most of whom did not identify their grandparents) (Figure 3). Curiously, Flegontov et al. (2016b) have not questioned the localization of Sephardic Jews, nor proposed an admixed origin for these populations predicted adjacently to AJs or overlapped with them. Overall, Flegontov et al.'s (2016b) speculation that AJs have experienced a recent admixture with Europeans is unsupported.

Question #11: Are AJs an unmixed, two-ways mixed, or highly mixed population?
Flegontov et al. (2016b) next asked whether the localization of AJs can be explained by mixture of European and Middle Eastern populations or proxies among modern-day populations. We confirmed the predicted location of AJs by showing the high similarity of AJ's admixture proportions to that of "native individuals" generated from GPS's genetic-geographic model (Das et al. 2016, Figure 5). In other words, the GPS model supports a scenario where Ashkenazic Jewish admixture proportions have occurred in "ancient Ashkenaz," primarily through Judaization of local populations. These local populations were probably the vast Greco-Roman Godfearers recorded living along the shores of the Black Sea (Baron 1937) at least since around 680 B.C. (Carpenter 1948). The predominant contribution of these Southern European populations to the Ashkenazic Jewish genome (60-80%) has been recently confirmed by Xue et al. (2016). However, as has been suggested by (Das et al. 2016), it is likely that a more complex admixture event involving several neighboring populations took place.

Question #12: What are the hypotheses concerning the origin of AJs?
Flegontov et al. (2016b) stated that "the traditional view on the history of the European Jewish diaspora: its Levantine origin, migration to the North Mediterranean followed by substantial local admixture, especially on the maternal side, and subsequent limited East European admixture in the Ashkenazi community" and cited in support of this five studies (Atzmon et al. 2010; Behar et al. 2010; Behar et al. 2013; Costa et al. 2013; Rootsi et al. 2013). This statement contains several inaccuracies: First, none of the studies Flegontov et al. cited support this assertion. Second, the contemporary debate on the origin of Ashkenazic Jews is largely captured by the two competing Rhineland and Irano-Turko-Slavic hypotheses (Table 1), neither of which depicts a "migration to the North Mediterranean". Third, Flegontov et al.'s (2016b) scenario consisted of a "substantial local admixture," which was earlier criticized by the authors (see Question #10). Fourth, Flegontov et al. (2016b)'s proposed migration scenario lacks any evidential support from history. Fifth, the Rhineland hypothesis is the traditional view, not the proposed scenario.

Question #13: What are the geographical origins of AJs reported in the literature?
Flegontov et al.'s (2016b) statement that the results of Behar et al.'s (2013) biogeographical analysis supports a Middle Eastern origin for AJs is incorrect. Thus far, three biogeographical analyses have ben published using three distinct approaches and largely different datasets (Behar et al. 2013, Figure 2b; Elhaik 2013, Figure 4; Das et al. 2016, Figure 4). All three analyses identified Turkey as the predominant origin of AJs. This finding is in support of the Irano-Turko-Slavic hypothesis (Figure 3) and at odds with the Rhineland hypothesis, which, if one entertains a more extreme view, consists of an epic tale composed of morbid exilic and supernatural events, none of which is supported by the data (Table 1). Remarkably, these recognizable and acknowledgeable limitations (e.g., Aptroot 2016) do not deter proponents of this hypothesis. The critics' argument that there is a linguistic support for this theory remains unconvincing and is unsuitable for debate in a genetic journal.

Question #14: Can evidence from ancient DNA or local ancestral analyses challenge the Turkish origin of AJs?
Flegontov et al. (2016b) hypothesized that alternative approaches to the question of AJ origin may yield different results from those reported in the literature (Figure 3). The authors remarked that "only studies of ancient genomes and their coordinates in space and time can approach locating ancestral homelands with enough precision" and advocated the use of "more data-intensive and sophisticated approaches for the study of population history within the last five thousand years" with tools like GLOBETROTTER (Hellenthal et al. 2014). Two recent studies allow this hypothesis to be tested.

In the ancient DNA analysis of six Natufians and a Levantine Neolithic (Lazaridis et al. 2016), some of the most likely Judaean progenitors (Finkelstein and Silberman 2002; Frendo 2004), the ancient individuals clustered predominantly with modern-day Palestinians and Bedouins and marginally overlapped with Arabian Jews. AJs clustered away from these ancient Levantine individuals and adjacent to Neolithic Anatolians and Late Neolithic and Bronze Age Europeans. AJs also clustered between Turkish and Italian Jews adjacently to south Italians. These findings are consistent with the predictions of the Irano-Turko-Slavic hypothesis ([Table 1](#)). In the second analysis, Xue et al. (2016) have applied GLOBETROTTER to a dataset of 2,540 AJs genotyped over 252,358 SNPs. The inferred ancestry profile for AJ was 5% Western Europe, 10% Eastern Europe, 30% Levant, and 55% Southern Europe. The authors believed that the Levant ancestry might be somewhat higher, although it is probably inflated due to the misclassification of Druze, a population of Near Eastern origin (Shlush et al. 2008; Elhaik 2013), as Middle Eastern. Elhaik (2013) reported similar Middle Eastern ancestral proportions (25-30%). The remaining ancestral proportions cannot be compared since Xue et al. (2016) ignored the Caucasus ancestry. The high Southern European ancestry reported by the authors can be explained by the presence of Greco-Roman populations in the Black Sea during the first century A.D and is consisted with Das et al.'s (2016) hypothesis that "Ashkenazic Jewish genomes may be conglomerates of Greco-Roman-Turko-Irano-Slavic and perhaps Judaean genomes formed through ongoing proselytization events that continued undisturbed for many centuries in Turkish "Ashkenaz"." Xue et al.'s (2016) inferred an "admixture time" (which, to the best of our understanding, corresponds to the time the admixture event occurred) of 960-1,416 AD (≈24-40 generations ago). This date corresponds to the time AJs experienced major geographical shifts as the Khazar kingdom diminished and their trading networks collapsed forcing them to relocate to Europe (Das et al. 2016). The lower boundary of that date corresponds to the time Slavic Yiddish originated, to the best of our knowledge.

Question #15: Is it possible that Yiddish was invented in Germany?
Flegontov et al. (2016b) cite linguistic evidence that allegedly supports a German origin for Yiddish, however this ignores the mechanics of relexification, the linguistic process which produced Yiddish and other "Old Jewish" languages (i.e., those created by the 9-10$^{th}$ century). Understanding how relexification operates is essential to understand the evolution of languages. This argument has a similar context to that of conservation of function in whales. Rejecting the theory of evolution may lead one to conclude that the Cetaceans are a clade of odd fishes. By disregarding the literature on relexification and Jewish history in the early Middle Ages, the authors reach conclusions that have weak historical support. The advantage of a biogeographical analysis is that it allows us to infer the geographical origin of the speakers of Yiddish, where they resided, and with whom they intermingled, independently of historical controversies, which provides a data driven view on the question of geographical origins. This allows objective review of potential linguistic influences on Yiddish (Table 1), which exposes the dangers in adopting a "linguistic creationism" view in linguistics.

Question #16: What is the historical evidence in favor of a Slavic origin?
Flegontov et al. (2016b) next asked for further evidence for the Slavic origin of Yiddish beyond the linguistic evidence proposed by Wexler. The authors also implied that the Silk Road played a minor role in the formation of Yiddish compared to German dialects.

The historical evidence is paramount. Jews played a major role on the Silk Roads in the $9^{th}$-$11^{th}$ century. In the mid-$9^{th}$ century, in roughly the same years, Jewish merchants in both Mainz and at Xi'an received special trading privileges from the Holy Roman Empire and the Tang dynasty court (Robert 2014). These were the very roads that linked Xi'an to Mainz and Andalusia, and further to sub-Saharan Africa and across to the Arabian Peninsula and India-Pakistan. The Silk Roads provided the motivation for Jewish settlement in Afro-Eurasia in the $9^{th}$-$11^{th}$ centuries since the Jews (sometimes along with "pagans," e.g., the Rus', the Kiev-Polissians who were to provide the basis for the contemporary Ukrainian people) played a dominant role on these routes as a neutral trading guild with no political agendas (Gil 1974; Cansdale 1996; Cansdale 1998). Hence, the Jewish traders had contact with a wealth of languages in the areas that they traversed (Hadj-Sadok 1949; Khordadhbeh 1967; Wexler TBD). The Silk Roads were controlled by Iranian polities (the Persians and the Sogdians), which provided opportunities for Iranian-speaking Jews, who constituted the overwhelming bulk of the world's Jews from the time of Christ to the $11^{th}$ century (Baron 1952). A Persian official in the Umayyad Caliphate in the $9^{th}$ century who met Iranian Jewish merchants lists six languages spoken by the Jewish merchants, but it can be shown that the list includes both individual and sets of languages (Gil 1974). Hence, the total number of languages spoken by the Jewish merchants could reach as many as a dozen. It should not come as a surprise to find that Yiddish (and other Old Jewish languages) contains components and rules from a large variety of languages, all of them spoken on the Silk Roads (Khordadhbeh 1967; Wexler 2011; Wexler 2012; Wexler TBD). Therefore, more attention should be given to Silk Road languages than to German dialects—which have made only a modest contribution to Yiddish structure.

In addition to language contacts, the Silk Roads also provide the motivation for widespread conversion to Judaism—e.g., by Iranians, Greeks, Slavs, Berbers, Arabians and Himyarites— all eager to participate in the extremely lucrative trade along the Silk Road, which had become a Jewish quasi-monopoly (Rabinowitz 1945; Rabinowitz 1948; Baron 1957). These conversions are discussed in Jewish literature between the $6^{th}$ and $11^{th}$ centuries, both in Europe and Iraq. Yiddish and other Old Jewish languages (e.g.,

Judezmo [Judeo-Spanish], Judeo-Arabic, Judeo-Georgian, Judeo-Berber) were all created by the peripatetic merchants as secret languages that would isolate them from their customers and non-Jewish trading partners (Gil 1974; Cansdale 1996; Cansdale 1998) (Hadj-Sadok 1949; Khordadhbeh 1967; Robert 2014). This means that the study of Yiddish genesis necessitates the study of all the Old Jewish languages of the period in question. It is difficult to expect students of Yiddish who train themselves in German, Slavic and Hebrew (at best) to fully comprehend the origin of Yiddish without understanding Iranian languages. Future research is necessary to understand whether association with Jewish merchants is the main factor which brought the nomadic Roma (Gypsies) out of India to Iran, Europe, and the Far East. The lexicon shared by Romani and Yiddish suggests that this was the case (e.g., Littmann 1920; Pstrusińska 1990; Pstrusińska 2004; Den Besten 2008).

Question #17: What is the validity of the relexification hypothesis?
Flegontov et al. (2016b) next asked whether the existence of Slavic elements in Yiddish could be explained by "missing inheritance," i.e., parents with imperfect knowledge in Yiddish passing on Slavic elements to their offspring, as opposed to the prediction of the relexification hypothesis. We first note that the history of the relexification hypothesis begun seven decades ago, having first been adopted by Creole studies (Faine 1939), although it has probably even deeper roots. Although relexified languages were identified as far back as the 17th century in Europe no special word was coined for them. For example, Buxtorf's dictionary (1645) separated genuine Hebraisms from pseudo-Hebraisms created in the process of relexification from Yiddish and other Jewish languages. Most of the early Jewish renditions of the Hebrew-Aramaic Bible into the colloquial languages of the Jews are not free translations but strict copies of Hebrew-Aramaic syntax, morphology, and lexicon/semantics—in other words, acts of relexification and not translation. Hence, a speaker of Judezmo cannot possibly understand the Ladino version of the Bible unless, of course, he is fluent in the underlying Biblical text. The proposal that Slavism penetrated Yiddish through mispronunciations of the parents is unsupported and inconsistent with the Jewish culture of scholarship from an early age at the hands of educated teachers.

Question #18: Is there a significant number Iranian or Turkic elements in Yiddish?
This question is preposterous given Jewish history. The Babylonian Talmud, completed by the 6th century A.D., is rich in Iranian linguistic, legalistic, and religious influences. From the Talmud, a large Iranian vocabulary has entered Hebrew and Judeo-Aramaic, and from there spread to Yiddish (e.g., Hebrew words *zman* 'time', *dat* 'religion', *pardes* 'orchard', *bdika* 'examination', *gniza* 'storage', *gizbar* 'treasurer' are of Iranian origin). This corpus has been known since the 1930s and is common knowledge to Talmud scholars (Telegdi 1933). It was Iranian Jews of heterogeneous ethnic origins who brought

Iranian Judaism to the Khazars, along with Iranian and Aramaic; in the Khazar Empire, the Turkic and Iranian Jews became speakers of Slavic—an important language because of the trading activities of the Rus' (pre-Ukrainians) with whom the Jews were undoubtedly allied on the routes linking Baghdad and Bavaria.

The authors claim that most of the Yiddish vocabulary (we can expand this to include syntax and phonology as well) is of German origin. However, we invite them to have a closer look at the Germanic component in Yiddish. First, we can "predict" in retrospect with almost total accuracy which German elements would be accommodated in Yiddish and which would be rejected. This is made possible by a comparison of Slavic and German structures and is a clear indication that Yiddish was invented by mapping German phonetic strings onto the substratal native languages of the speakers—Slavic, Iranian and Turkic (Wexler 1991; Wexler 2002). Second, probably half, if not more, of the "German" components in Yiddish are not comprehensible to any native speaker of German. For example, Yiddish *unterkojfn* 'to bribe' means nothing to a German, who would of course recognize German *unter-* 'under' and *kaufen* 'to buy', but the combination is ungrammatical in that language—though not in Slavic, which licensed the creation of the "Germanism" in Yiddish (Wexler 1991). There are thousands of such examples (Wexler 1991; Wexler 2002; Wexler 2011; Wexler 2012). German expressions which violate Slavic syntactic and morphological parameters will be unlicensed for use in Yiddish during the relexification process.

**Conclusions**

Producing accurate geographical predictions, with a resolution down to home village in humans, remains the ultimate goal of biogeography. Elhaik et al. (2014) demonstrated that the Geographic Population Structure (GPS) framework allows, in some cases, reaching such levels of accuracy. Two recent studies have applied GPS to populations genotyped on the Genochip microarray to shed light on open questions in genetics and linguistics: Flegontov et al. have (2016a) studied the origin of Siberian Kets and their language, and Das et al. (2016) studied the origin of AJ and their language. The increasing usage of GPS to study ancestries and languages deeply rooted in the past is intriguing, but caution is warranted in interpreting the findings. We also note that more evolutionary understanding should be implemented in linguistics. That includes giving more attention to the linguistic process that alter languages (e.g., relexification) and acquiring more competence in other languages and histories. When studying the origin of Ashkenazic Jews and Yiddish, such knowledge should include the history of the Silk Roads and Irano-Turkish languages. We hope that this perspective would be useful to understanding the strengths and limitations of the GPS framework and how it can best be applied to answer historical and linguistic questions.

**Competing interests**

EE is a consultant to DNA Diagnostic Centre.

**Acknowledgment**

E.E was partially supported by a Genographic grant (GP 01-12), The Royal Society International Exchanges Award to E.E. and Michael Neely (IE140020), MRC Confidence in Concept Scheme award 2014-University of Sheffield to E.E. (Ref: MC_PC_14115), and a National Science Foundation grant DEB-1456634 to Tatiana Tatarinova and E.E.

**Tables**

Table 1
**Major open questions regarding the origin of AJs and Yiddish language as explained by two competing hypotheses.** The genetic evidence produced by Das et al. (2016) is shown in the last column.

| Open questions | Rhineland hypothesis | Irano-Turko-Slavic hypothesis | Evidence produced by Das et al. (2016) |
| --- | --- | --- | --- |
| The term "Ashkenaz" | Used in Hebrew and Yiddish sources from the 11[th] century onward to denote a region in what is now roughly Southern Germany (Wexler 1991; Aptroot 2016). | Denotes an Iranian people "near Armenia," presumably Scythians known as *aškuza, ašguza,* or *išguza* in Assyrian inscriptions of the early 7[th] century B.C. (Wexler 2012; Wexler 2016). | GPS analysis uncovered four primeval villages in northeastern Turkey whose names resemble "Ashkenaz," at least one of which predates the Jewish settlement in Germany. "Ashkenaz" is thereby a placename associated with this region and its populations. |
| The ancestral origin of Ashkenazic Jews | Judaean living in Judaea until 70 A.D. who were exiled by the Romans (King 2001). This scenario has no historical (Sand 2009) nor genetic support (Elhaik 2013; Lazaridis et al. 2016). | A minority of Judaean emigrants and Irano-Turko-Slavic converts to Judaism (Wexler 2012). | The admixture proportions of Ashkenazic Jews is affiliated with northeastern Turkey and is similar to those of local and neighboring Jewish and non-Jewish populations. |

| The arrival of Jews to German lands | The Romans exiled the Palestinian Jews (70 A.D.) to Roman lands. Jewish merchants and soldiers arrived to German lands with the Roman army and settled there. (King 2001). This scenario has no historical support (Wexler 1993). | Jews from the Khazar Empire and the former Iranian Empire plying the Silk Roads began to settle in the mixed Germano-Sorbian lands during the first Millennium (Wexler 2011). | The admixture proportions of Ashkenazic Jews were predicted to a Near Eastern hub of ancient trade routes that connected Europe, Asia, and the northern Caucasus. The findings suggest that migration to Europe took place initially through trade routes and later through Khazar lands. |
|---|---|---|---|
| Yiddish's emergence in the 9th century | Between the 9th and 10th centuries, French- and Italian-speaking Jewish immigrants adopted and adapted the local German dialects (Weinreich 2008). | Upon arrival to German lands, Western and Eastern Slavic went through a relexification to German, creating what became known as Yiddish (Wexler 2012). | |

| Growth of Eastern European Jewry | A small group of German Jews migrated to Eastern Europe and reproduced via a so-called "demographic miracle," which resulted in an unnatural growth rate (1.7-2% annually) (van Straten and Snel 2006; van Straten 2007) over half a millennia acting only on Jews residing in Eastern Europe (Ben-Sasson 1976; Atzmon et al. 2010; Ostrer 2012). This explanation is indubitably fictitious. | During the half millennium (740–1250 CE), Khazar and Iranian lands harbored the largest Eurasian Jewish centers. Ashkenazic, Khazar, and Iranian Jews then sent offshoots into the Slavic lands (Baron 1957). | Most of the Ashkenazic Jews (93%) were predicted to Northeastern Turkey and the remaining individuals clustered along a gradient ending in Eastern European lands. The German origin of Jews is unsupported by the data. |

Table 2

**A summary of the questions raised by Flegontov et al.**

| # | Questions | Brief answers |
|---|---|---|
| 1 | How should GPS predictions be interpreted? | GPS traces unmixed individuals to the region where populations with the most similar admixture proportions to those of the individual are found. This is the place where the final admixture event occurred at the population level, i.e., the last massive demographic event that changes the allele frequencies of the entire population. |
| 2 | How does GPS behave in the case of two-way mixed individuals? | GPS was designed to handle unmixed individuals. Two-ways admixed individuals would be predicted to the middle point of their parental populations, as both are "pulling" the in equal strengths (Figure 1). The next GPS version would position such individuals in their parental countries. |
| 3 | Is tracing a population movement back in time feasible? | No. GPS cannot trace movements. |
| 4 | Does the choice of Genochip markers bias GPS predictions? | No. The Genochip microarray consists of ~130,000 common and rare alleles ancestry informative markers. Elhaik et al. (2014) showed that it yields highly accurate predictions. Whole-genome data remain expensive and are available for a small number of populations. |
| 5 | Did results from an older study prompt the localization in a later study? | No. These were two separate and unrelated trials. |

| 6 | Was the localization of Sardinians to their villages due to noise? | No. The noise in GPS is very small, which allows accurate localizations to be made based on such small admixture proportions. The differences in admixture proportions between villagers are expected to be very small and are not noise, which distributes randomly, and thereby increases the genetic distances between neighboring villages. |
|---|---|---|
| 7 | Does the absence of reference populations reduce the accuracy of the results? | Yes. GPS relies on reference population to "pull" samples. A comprehensive coverage, particularly in island and coast-line populations is necessary to derive accurate predictions. |
| 8 | Can inaccurate localizations be due to shared ancestry or inaccurate modelling? | Shared ancestry. Incorrect localizations due to shared ancestry would usually predict individual to neighboring countries, whereas the alternative would yield random predictions. We did not observe random predictions. |
| 9 | Why are GPS predictions incorrect for some of the non-Jewish populations in Das et al. (2016)? | These predictions were made with the leave-one-out population approach. This is a highly restrictive approach adopted to estimate the error in locating populations that are absent from the reference panel. |
| 10 | Can we infer the likelihood of European admixture in AJs from the demographic data? | No. |
| 11 | Are AJs an unmixed, two-ways mixed, or highly mixed population? | GPS predictions fit well with AJs emerging from a local population in today's Turkey, however a more complex admixture history is likely. |
| 12 | What are the hypotheses concerning the origin of AJs? | The two major hypotheses are summarized in Table 1. |
| 13 | What are the geographic origins of AJs reported in the literature? | All biogeographical studies reported a predominant Turkish origin (Figure 3). |

| 14 | Can evidence from ancient DNA or local ancestral analyses challenge the Turkish origin of AJs? | No. Two such studies yield similar results to those reported by Das et al. (2016). |
| --- | --- | --- |
| 15 | Is it possible that Yiddish was invented in Germany? | No. Such proposal ignores Jewish history, genetic evidence, and the evolution of languages (Table 1). |
| 16 | What is the historical evidence in favor of a Slavic origin? | Jews have played a major role on the Silk Roads in the 9th-11th century. In addition to language contacts, the Silk Roads also provide the motivation for widespread conversion to Judaism to populations eager to participate in the extremely lucrative trade, which had become a Jewish quasi-monopoly. This necessitated developing a secret language to maintain this monopoly. |
| 17 | What is the validity of the relexification hypothesis? | The history of the relexification hypothesis begun at least seven decades ago. Relexification is the key to understanding Jewish languages because most of the early Jewish renditions of the Hebrew-Aramaic Bible into the colloquial languages of the Jews are acts of relexification and not translation. |
| 18 | Is there a significant number Iranian or Turkic elements in Yiddish? | Yes. The Babylonian Talmud is rich in Iranian vocabulary, which entered Hebrew and Judeo-Aramaic, and from there to Yiddish. |

**Figure legend**

Figure 1

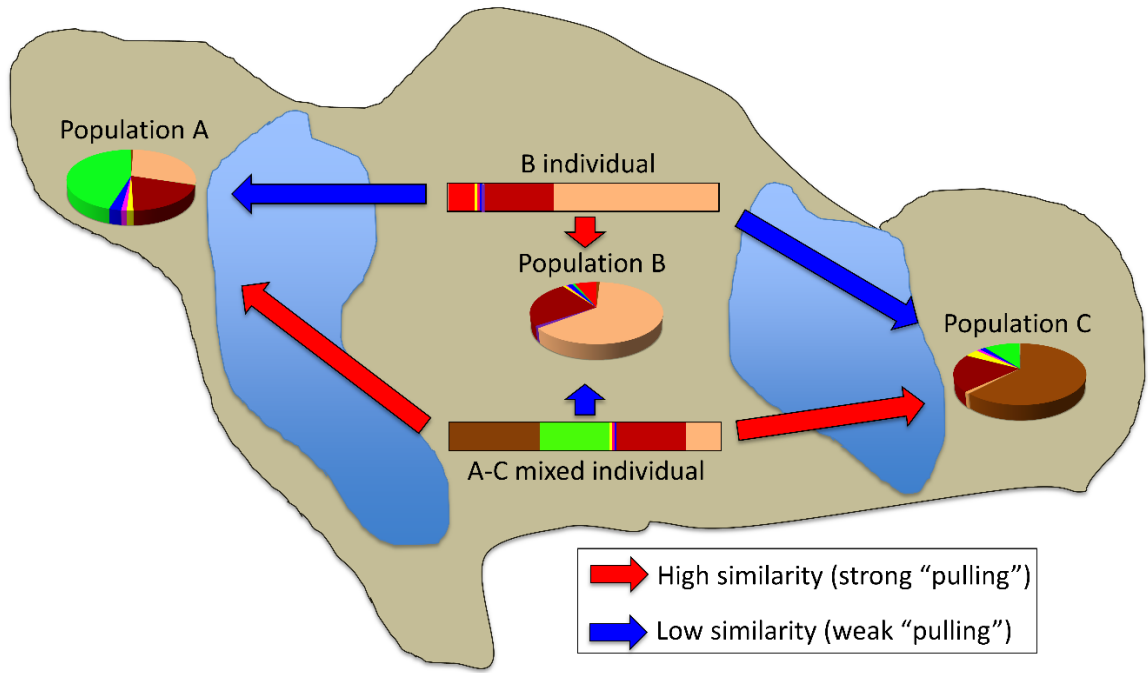

**Illustration of GPS localization model for unmixed and mixed individuals**. In determining the location of unmixed individual B, the individual's admixture proportions are compared to those of three reference populations (A, B, and C). The genetic distances between individual B and populations A and C are high, thereby their "pull" is weak and their effect on the final location of this individual is minor, compared to that of the true parental population B. A-C mixed individual is predicted incorrectly to the region of population B, which happened, by chance, to reside between populations A and C, both of which are "pulling" the individual in equal strengths. Evidently, B is not A-C's parental population since their admixture proportions are very different.

Figure 2

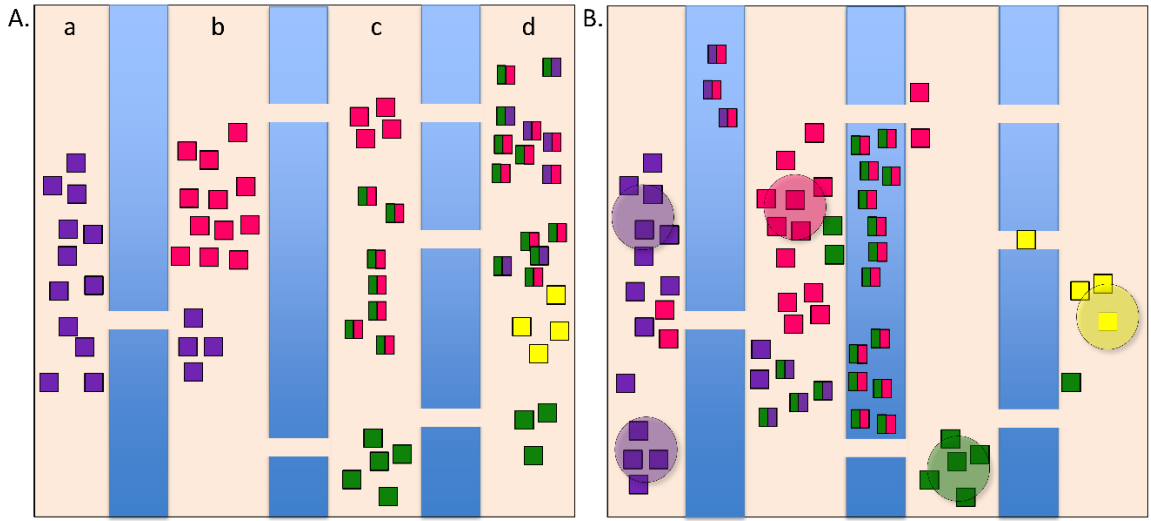

**An illustration of GPS results**. A hypothetical world consists of four regions (a-d) that vary in the degree of isolation due to natural barriers. Descendants of four unmixed populations are shown by single-color squares alongside two-ways admixed individuals shown by color-matched squares. The modern-day residency of individuals is shown in A. GPS predictions (B) are made using a panel of four reference populations (circles) positioned in the ancestral locations of the unmixed populations that gravitate genetically similar individuals towards them. GPS predicts most of the unmixed individuals to the ancestral location of their population with some inaccuracies due to the shared history of neighboring populations. The mixed individuals are predicted incorrectly to the region between their parental populations.

Figure 3

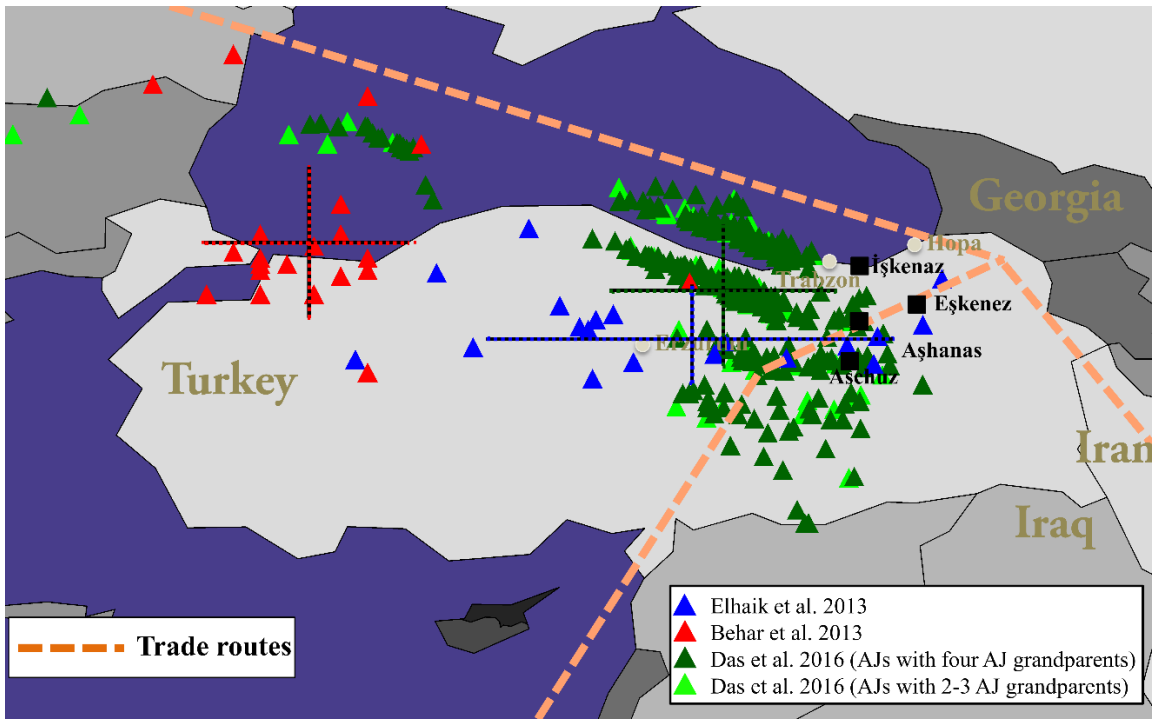

**Biogeographical localization of AJs based on three studies**. Geographical predictions of individuals analyzed in three separate studies employing different tools: Elhaik (2013, Figure 4) (blue), Behar et al. (2013, Figure 2b) (red), and Das et al. (2016, Figure 4) (dark green for AJs who have four AJ grandparents and light green for the rest) are shown. Color matching mean and standard deviation (bars) of the longitude and latitude are shown for each cohort. Since we were unsuccessful in obtaining the data points of Behar et al. (2013, Figure 2b) from the corresponding author, we procured 78% of the data points from their figure. Due to the low quality of the figure we were unable to reliably extract the remaining data points.